\documentclass{webofc}
\usepackage[varg]{txfonts}   
\usepackage{soul}

\begin{document}

\title{Graph Neural Network for Object Reconstruction \\ in Liquid Argon Time Projection Chambers}

\author{\firstname{V} \lastname{Hewes}\inst{1}\fnsep\thanks{\email{vhewes@fnal.gov}} \and
        \firstname{Adam} \lastname{Aurisano}\inst{1}\fnsep \and
        \firstname{Giuseppe} \lastname{Cerati}\inst{2}\fnsep \and
        \firstname{Jim} \lastname{Kowalkowski}\inst{2}\fnsep \and
        \firstname{Claire} \lastname{Lee}\inst{3}\fnsep \and
        \firstname{Wei-keng} \lastname{Liao}\inst{3} \and
        \firstname{Alexandra} \lastname{Day}\inst{3} \and
        \firstname{Ankit} \lastname{Agrawal} \inst{3} \and
        \firstname{Maria} \lastname{Spiropulu}\inst{4} \and
        \firstname{Jean-Roch} \lastname{Vlimant}\inst{4} \and
        \firstname{Lindsey} \lastname{Gray}\inst{2} \and
        \firstname{Thomas} \lastname{Klijnsma}\inst{2} \and
        \firstname{Paolo} \lastname{Calafiura}\inst{5} \and
        \firstname{Sean} \lastname{Conlon}\inst{5} \and
        \firstname{Steve} \lastname{Farrell}\inst{5} \and
        \firstname{Xiangyang} \lastname{Ju}\inst{5} \and
        \firstname{Daniel} \lastname{Murnane}\inst{5}
}

\institute{University of Cincinnati, Cincinnati, OH, USA
\and
           Fermilab, Batavia, IL, USA
\and
           Northwestern University, Evanston, IL, USA
\and
           California Institute of Technology, Pasadena, CA, USA
\and
           Lawrence Berkeley National Laboratory, Berkeley, CA, USA
          }

\abstract{%
  This paper presents a graph neural network (GNN) technique for low-level reconstruction of neutrino interactions in a Liquid Argon Time Projection Chamber (LArTPC). GNNs are still a relatively novel technique, and have shown great promise for similar reconstruction tasks in the LHC. In this paper, a multihead attention message passing network is used to classify the relationship between detector hits by labelling graph edges, determining whether hits were produced by the same underlying particle, and if so, the particle type. The trained model is 84\% accurate overall, and performs best on the EM shower and muon track classes. The model's strengths and weaknesses are discussed, and plans for developing this technique further are summarised.
}

\maketitle

\section{Introduction}
\label{sec:intro}

The next generation of neutrino physics experiments heavily utilises LArTPC detectors, which measure particle interactions with precise spatial resolution. The wealth of rich information provided by a LArTPC means automated particle reconstruction can prove challenging, and in recent years machine learning techniques have been increasingly adopted to meet this challenge due to their ability to outperform traditional methods \cite{dune-cvn}.

Meanwhile, in the context of the LHC, the Exa.TrkX collaboration has demonstrated that GNN-based methods show great promise for reconstructing detector hits into particle tracks \cite{exatrkx}. Such methods have the advantage of operating on any data structure which can be described by quantised nodes and their relationships, and also on heterogeneous node definitions, which makes them much more broadly applicable than CNNs, which require input tensors arranged in a regular grid structure.

\begin{figure}
    \centering
    \includegraphics[width=\textwidth]{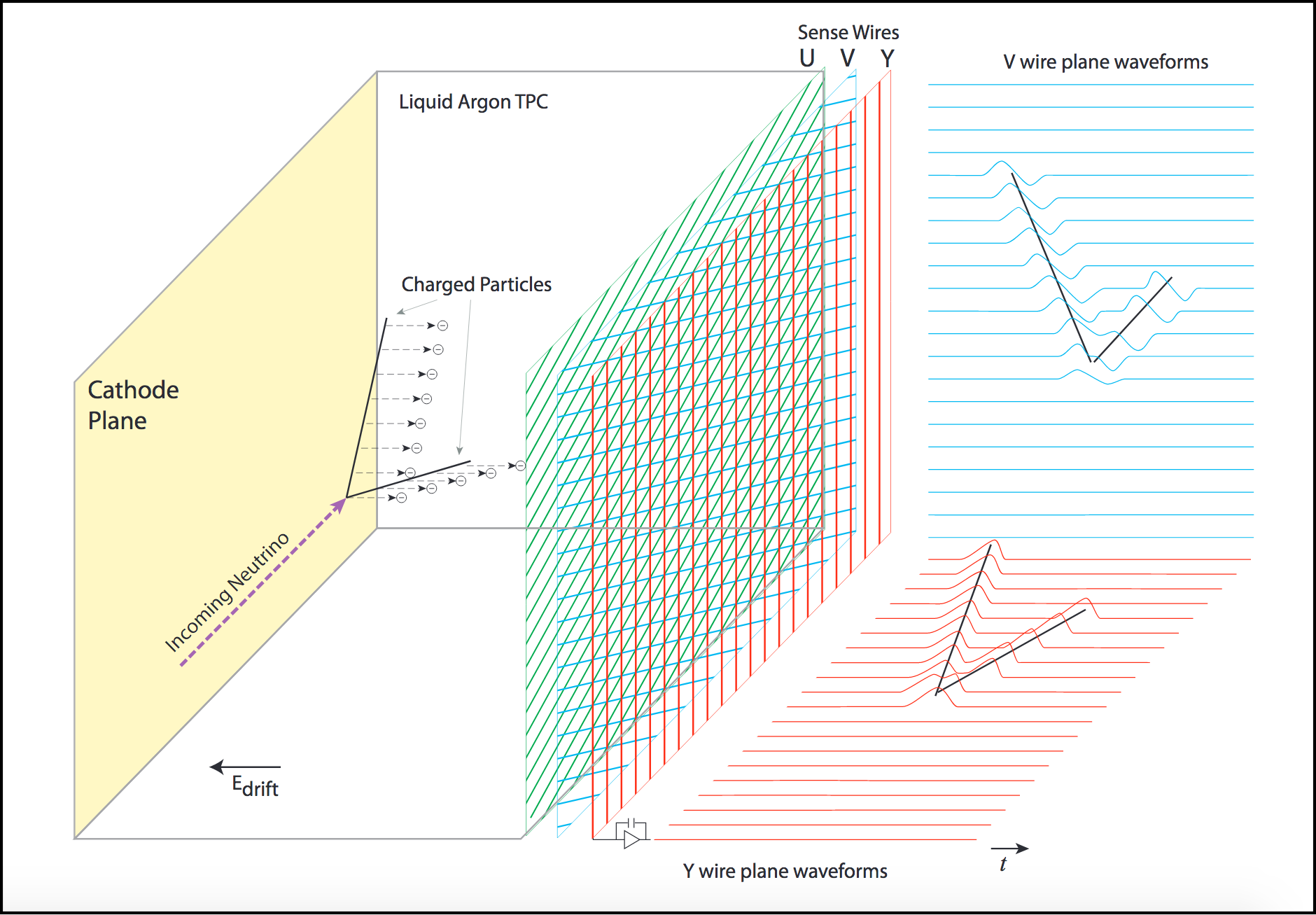}
    \caption{The operating procedure of a LArTPC -- ionisation electrons produced by charged particles in the TPC drift towards the collection planes under a strong electric field, and induce signals on the wire planes.}
    \label{fig:lartpc}
\end{figure}

The basic operating principles of a LArTPC are as follows: a cryostat filled with purified, cryogenically cooled liquid argon, with a cathode plane on one side, and three planes of wires (the ``anode plane assembly'', or APA) on the opposite side. Charged particles propagating through the inert argon will leave behind a trail of ionisation electrons. The cathode plane is held at a high voltage, inducing an electric field between the cathode and APA which causes these ionisation electrons to drift towards the APA. Upon reaching the APA, the ionisation electrons will induce a signal on the first two wire planes (the U and V planes, or ``induction planes'') before being collected by the final wire plane (the Y plane, or ``collection plane''). Due to the varying pitch angles of the three wire planes, each wire plane gives a different 2D representation of the interaction, which can be reconstructed back to a 3D representation. This process is demonstrated in Fig.~\ref{fig:lartpc}.

LArTPCs are a detector technology seeing heavy use in contemporary neutrino physics, both currently operating (ICARUS \cite{icarus}, MicroBooNE \cite{uboone}) and in construction (SBND, DUNE \cite{dunetdr}). LArTPCs produce high-resolution outputs which are globally sparse and often very large. Since neutrino interactions typically have high sparsity, recent studies have shown considerable efficiency improvements by moving to sparse CNNs \cite{sparse}, but even in these techniques, native detector outputs must be transformed through voxelization in order to produce CNN-compatible inputs. If efficient GNN reconstruction techniques can be developed, they will be able to operate on detector output without any form of downsampling.

\section{Methodology}

\subsection{Graph construction}
\label{sec:graph}

Simulated charged current quasielastic (CCQE) $\nu_{\mu}$ and $\nu_{e}$ interactions in a LArTPC geometry are used to construct 2D detector hit graphs. Neutrino interactions are simulated using GENIE \cite{genie}; particles propagating inside the detector are then simulated with Geant4 \cite{geant}, and processed into wire waveforms by a detector simulation. The raw waveforms extracted from the detector undergo deconvolution \cite{deconv1,deconv2} and Gaussian hit-finding \cite{hitfinding} to produce discrete hit objects. The hits on each wire plane form the nodes of a graph -- each wire plane is treated independently, so three independent graphs are produced for each neutrino interaction.

This simulated dataset consists of 254390 graphs, split into a 95\% sample for training (241670 graphs), and a 5\% sample for validation and testing (12720 graphs). Graphs contain an average of 1529 nodes and 13348 edges. Since EM shower and muon nodes are unique to $\nu_{e}$ and $\nu_{\mu}$ graphs respectively, there are an average of 1097 EM shower nodes per $\nu_{e}$ graph and 1886 muon nodes per $\nu_{\mu}$ graph; there is an average of 68 hadronic nodes overall per graph, since this class occurs in graphs of both neutrino flavours. There are an average of 861 false and 238 hadronic edges per graph; likewise, there is an average of 14860 EM shower edges per $\nu_{e}$ graph, and 8638 muon edges per $\nu_{\mu}$ graph.

Each graph node's features comprise 7 spatial coordinates -- plane, wire and time coordinates in both the local coordinate system within each detector module and the global coordinate system across all modules, and the ID of the TPC in which the hit occurred -- as well as the integral and root mean square width of each hit. Edges are drawn between hits which occupy the same local region (5 wires and 50 time ticks); this window was set by hand to be sufficiently broad that distinct but adjacent energy depositions remain connected, but sufficiently narrow that the number of graph edges remains computationally manageable.

\subsection{Label construction}
\label{sec:label}

CCQE interactions consist of a primary charged lepton ($e$ or $\mu$) emerging from the neutrino interaction vertex, in addition to hadronic activity due to nuclear recoil (typically short, highly ionising proton tracks or diffuse energy deposition from neutrons).  All graph edges are placed into one of four categories: \textbf{shower-like}, \textbf{muonic}, \textbf{hadronic} and \textbf{false}.

Each simulated interaction contains a single charged lepton in the final state, in addition to some hadronic activity. Each simulated hit is traced back to the simulated particle responsible for producing it, and that particle's parentage is traced back to the particles produced in the neutrino interaction. Any hits whose ancestor is the primary lepton is tagged as \textbf{shower-like} or \textbf{muonic}, depending on lepton flavour, while the remainder are tagged as \textbf{hadronic}. Edge labels are then produced by matching these labels between hits -- if two hits were produced by the same simulated particle, their connecting edge is labelled according to the hit label; otherwise, the edge is labelled \textbf{false}. EM showers are considered dense objects, meaning hits produced by particles in the EM cascade are considered to have been produced by the particle that instigated the shower for the purpose of labelling.

Muon and hadronic edges tend to leave long straight tracks in the detector -- since protons are more highly ionizing than muons, muon tracks tend to be longer with less energy deposited per hit. Electrons will initiate an electromagnetic shower, which appears in the detector as a fuzzy cone of energy deposition emerging from the vertex. False edges typically appear at interaction vertices, where particles of different types interface with each other. It should be noted that under the current ground truth definition, Michel electrons at the end of a muon track are assigned the muon label, although in future iterations of the model they will be assigned a unique label.

\subsection{Model construction}
\label{sec:model}

An attention message-passing model is used to classify graph edges, derived from HEP.TrkX work in track reconstruction using GNNs at the LHC \cite{heptrkx}. This model takes inital features on each graph node, and then forms edge features by concatenating the incoming and outgoing node's features for each edge and performing convolutions to produce an attention score. This score is then used to propagate information across graph edges to form new node features; this procedure is repeated multiple times to allow information to flow across the graph.

The HEP.TrkX model was modified from a binary edge classifier to a multi-head self-attention model \cite{multihead}. Under this construction, the initial features on each node are repeated to form an independent set of features for each class. During the edge convolution stage, a separate attention score is produced for each class; instead of the sigmoid layer utilised in a binary attention classifier, a softmax layer is applied to normalise the sum of the class's edge attention scores to unity. If one class's attention score for a given edge is high, the scores for the other edges will be weighted down to compensate. These attention scores are used to propagate information and produce new node features independently for each class.

\begin{table}[]
    \centering
    \begin{tabular}{|c|c|}
        \hline
        Class & Weight \\
        \hline
        False & 3.87 \\
        EM Shower & 0.38 \\
        Muon & 0.94 \\
        Hadronic & 13.99 \\
        \hline
    \end{tabular}
    \caption{Class weights used to reweight the loss function. Each class's weight is inversely proportional to the frequency with which the class appears in the training set.}
    \label{tab:weights}
\end{table}

The model's objective function is a standard categorical cross-entropy loss between the true and reconstructed edge labels. Since some classes are much more highly represented in the training set than others, each class's contribution to the loss function is weighted proportionally to the inverse of how many edges hold that label in truth. Class weights are shown in Tab.~\ref{tab:weights}. The optimizer and hyperparameters used for training are shown in Tab.~\ref{tab:hyperparams} -- in addition, a learning rate policy is applied to reduce the learning rate by half if the loss does not improve for 5 epochs. Batch size is limited by the variation in graph size; the impact this has on computational efficiency is discussed in Sec.~\ref{sec:discussion}.

\begin{table}[]
    \centering
    \begin{tabular}{|c|c|}
        \hline
        Optimizer & AdamW \\
        Learning rate & 1e-4 \\
        Batch size & 4 \\
        \hline
    \end{tabular}
    \caption{Optimizer and hyperparameters used during model training.}
    \label{tab:hyperparams}
\end{table}

\section{Training results}

\begin{figure}
    \centering
    \includegraphics[width=\textwidth]{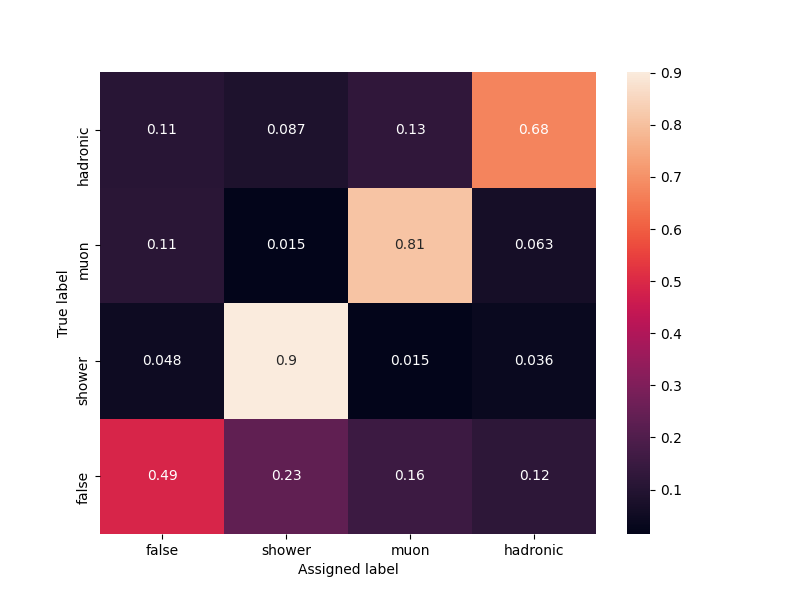}
    \caption{Confusion matrix showing the overlap of true and reconstructed edge labels.}
    \label{fig:confusion}
\end{figure}

The model was trained on an NVIDIA Tesla V100 GPU for 100 epochs. After reducing the class probability output to a single class score using an argmax operation, the model achieves 84\% accuracy when classifying graph edges, with a confusion matrix between classes shown in Fig.~\ref{fig:confusion}. The model achieves 90\% efficiency when identifying shower edges, and 81\% efficiency when identifying muon edges, but performs worse on the hadronic and false classes (68\% and 49\% respectively), which are less well represented in the training set.

\begin{figure}
    \centering
    \includegraphics[width=0.48\textwidth]{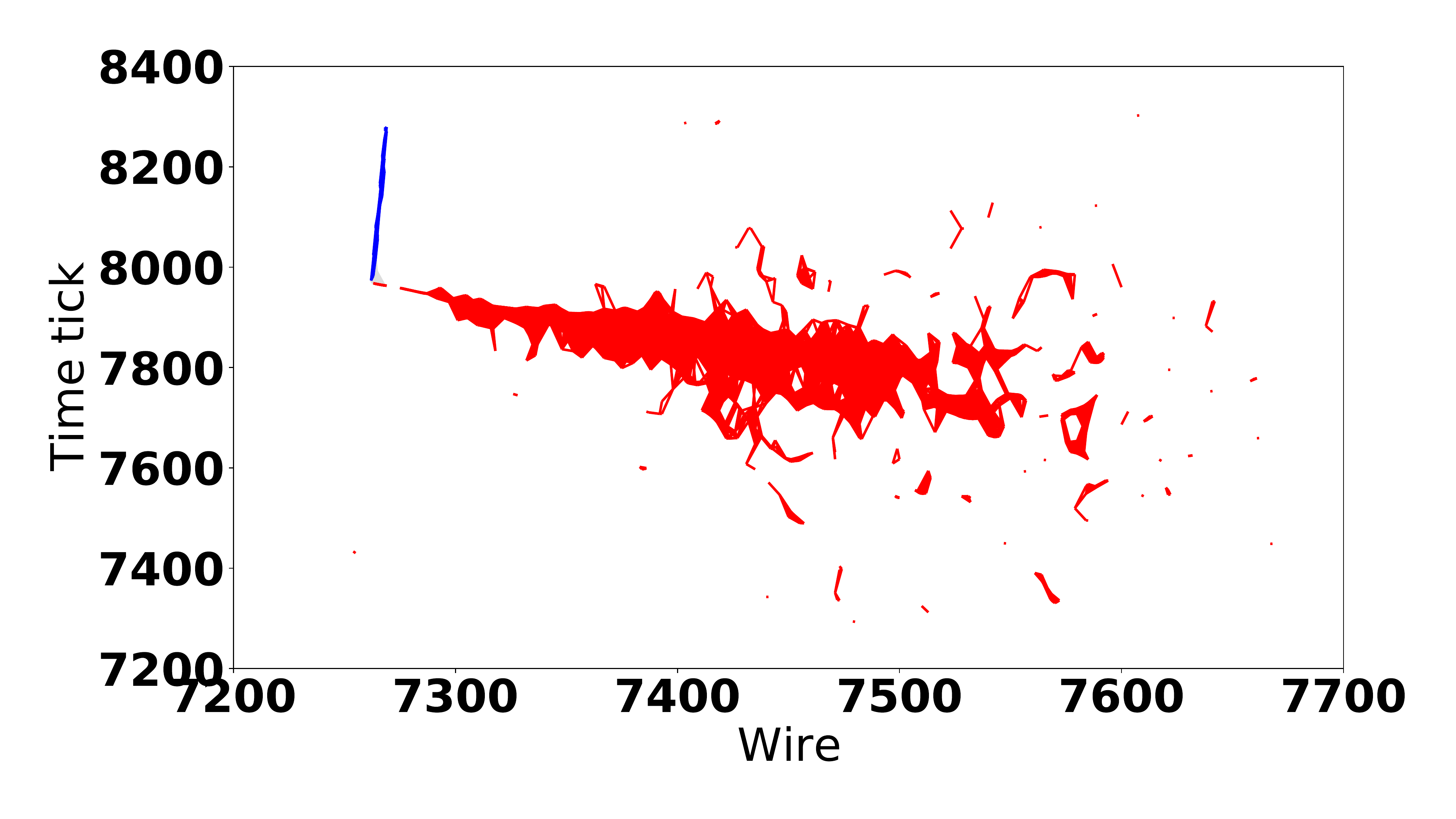}
    \hspace{0.01\textwidth}
    \includegraphics[width=0.48\textwidth]{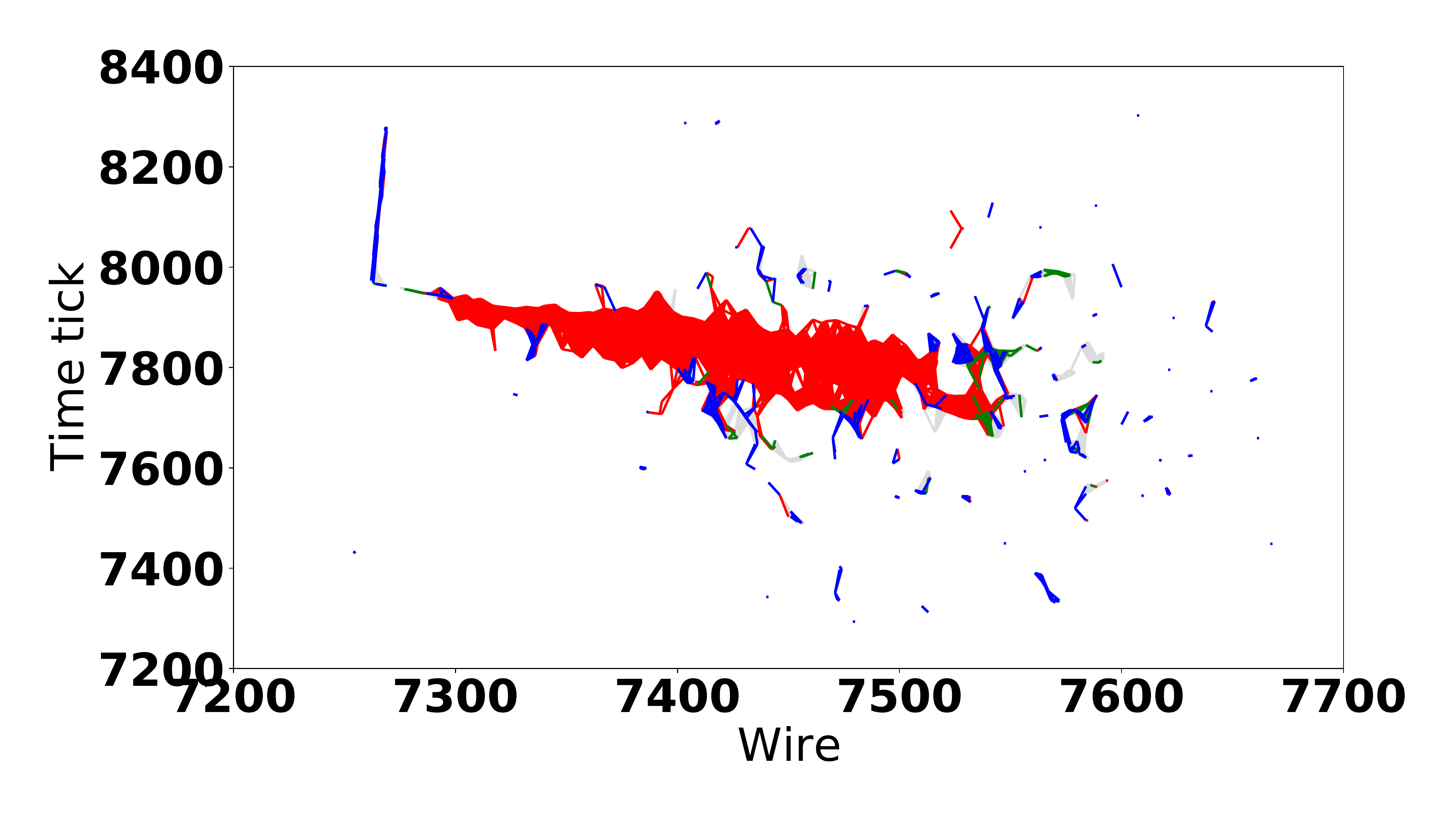}
    \caption{Example graph of a $\nu_{e}$ interaction (left: ground truth, right: model output). Shower-like edges are drawn in red, hadronic edges are drawn in blue, muonic edges are drawn in green and false edges are drawn in grey.}
    \label{fig:nue}
\end{figure}

Event displays comparing ground truth to model output for a $\nu_{e}$ interaction are shown in Fig.~\ref{fig:nue} -- the model correctly identifies the proton track in blue, and reconstructs the root of the shower, but tends to misclassify edges close to the shower boundary as being track-like.

\begin{figure}
    \centering
    \includegraphics[width=0.48\textwidth]{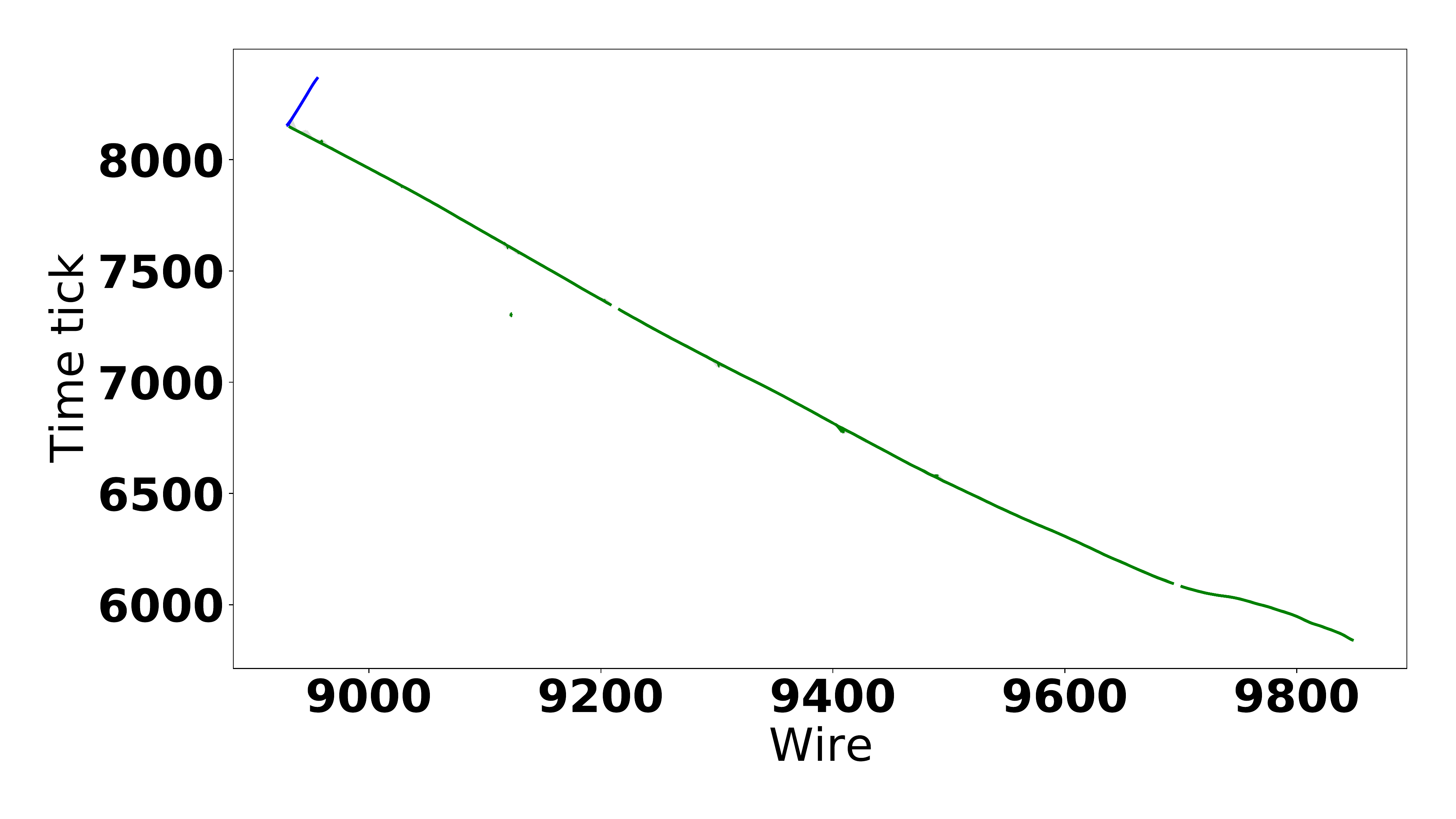}
    \hspace{0.01\textwidth}
    \includegraphics[width=0.48\textwidth]{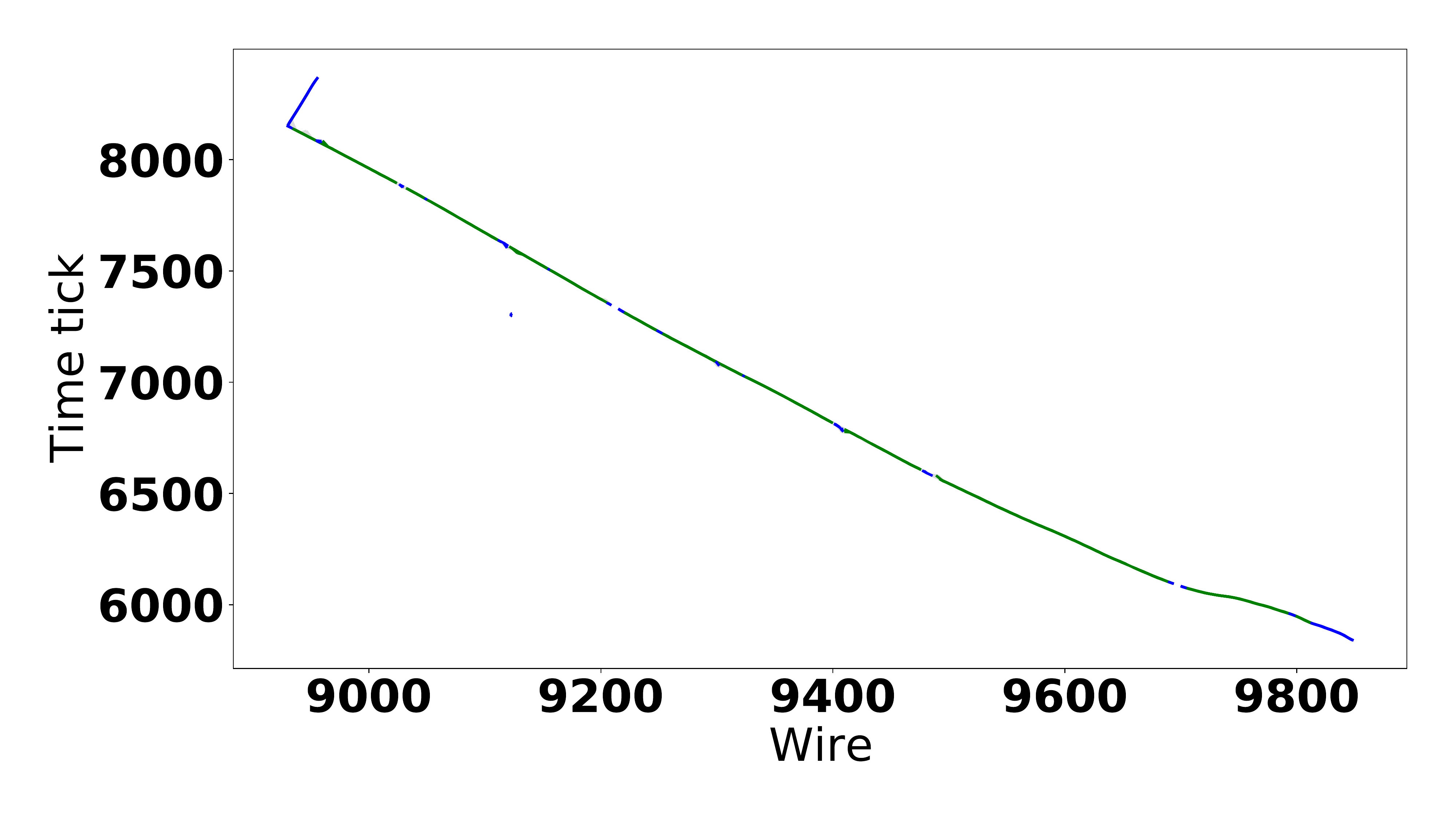}
    \caption{Example graph of a $\nu_{\mu}$ interaction (left: ground truth, right: model output). Shower-like edges are drawn in red, hadronic edges are drawn in blue, muonic edges are drawn in green and false edges are drawn in grey.}
    \label{fig:numu}
\end{figure}

Similarly, event displays for a $\nu_{\mu}$ interaction are shown in Fig.~\ref{fig:numu}. The model can typically identify edges as track-like, but can struggle to disambiguate the muon and hadronic labels, as seen in this example at the end of the muon track. In this example the model likely misclassifies the end of the muon track due to its Bragg peak, which increases energy deposition and can make the track segment look more proton-like. More rarely, the model can misclassify entire $\nu_{\mu}$ events, misidentifying the proton track as muon-like and the muon track as hadronic.

\begin{table}[]
    \centering
    \begin{tabular}{|c|c|}
    \hline
        Hidden features & 64 \\
        Message-passing iterations & 4 \\
        Train time per epoch & 9.5e3~s \\
        Validation time per epoch & 78~s \\
        GPU Memory & 14.3~GB \\ \hline
    \end{tabular}
    \caption{Model configuration, training time and memory usage.}
    \label{tab:model}
\end{table}

\section{Discussion}
\label{sec:discussion}

The attention message-passing network developed for use at the LHC forms the basis of the LArTPC model, but differences in detector technology pose unique problems that require more customised solutions solutions. The sequentially layered construction of the HL-LHC provides a natural constraint on the number of graph edges, as each hit on a given layer can only be connected to hits on the subsequent layer. The dense, monolithic nature of a LArTPC provides no such constraint, and so ad hoc constraints must be constructed to prevent the number of graph edges from becoming unmanageably large. As discussed in Sec.~\ref{sec:graph}, a simple region of interest based constraint is used in the work presented here, although more sophisticated techniques are in development.

Additionally, this model could be made more memory-efficient. The number of nodes in each event's graph depends on the spatial extent of the interaction. If several graphs with significantly more nodes than the mean are batched together, memory consumption will spike -- batch size must be limited to prevent these memory spikes from crashing training, but this means the memory consumption for the average batch is sub-optimal. A more sophisticated pseudo-random batching method which prevents large graphs from being batched together could reduce the model's memory overhead, and enable training with larger batch sizes.

Following the promising results of this GNN model, ongoing work will seek to expand and refine this method. Time-matching hits across wire planes will enable new graph edges that allow information to flow between the different 2D representations, leveraging this additional information to build a graph classification which is consistent between wire planes.

Additionally, utilising concepts from instance segmentation may allow misclassified outlier edges to be refined, such as graph edges on the outskirts of an EM shower misclassified as track-like, or proton track edges misclassified as muon-like and vice-versa. If the shower object can be identified as a coherent instance with a majority of shower-like edges, then edges at the shower boundary can be assigned the shower label even if they were individually misclassified by the network.

\section*{Acknowledgements}

This research was supported in part by the Office of Science, Office of High Energy Physics, of the US Department of Energy under Contracts No. DE-AC02-05CH11231 (CompHEP Exa.TrkX) and No. DE-AC02-07CH11359 (FNAL LDRD 2019.017).

This research used resources of the National Energy Research Scientific Computing Center (NERSC), a U.S. Department of Energy Office of Science User Facility located at Lawrence Berkeley National Laboratory, operated under Contract No. DE-AC02-05CH11231.

\end{document}